\documentclass[aps,floatfix,superscriptaddress,long,notitlepage,letterpaper,balancelastpage,nofootinbib,prd,floatfix,twocolumn]{revtex4}
\pdfoutput=1 

\usepackage{amsmath,mathtools,amssymb,amsthm,amsxtra,overpic,bbm,epsfig,subfigure,url,bm}
\usepackage{hyperref}
\usepackage{mathrsfs}
\usepackage{color,xcolor}
\usepackage{comment}
\usepackage{float}
\usepackage{enumitem}
\usepackage{slashed}

\usepackage{amsmath}

\definecolor{MyDarkBlue}{rgb}{0.1, 0.1, 0.8} 
\definecolor{SBlue}{rgb}{0.2, 0.4, 0.7} 
\definecolor{MyLightBlue}{rgb}{0.22,0.51,0.9}
\definecolor{MyGreen}{rgb}{0.0, 0.5, 0.0}
\definecolor{BrickRed}{rgb}{0.8, 0.25, 0.33}
\definecolor{nicered}{rgb}{0.7,0.12,0.12}
\definecolor{nicegreen}{rgb}{0.,0.5,0.}
\definecolor{niceblue}{rgb}{0.,0.,0.8}
\hypersetup{
colorlinks=true,
linkcolor=black,
filecolor=nicegreen,      
urlcolor=SBlue,
citecolor=SBlue,
}

\begin{document}

\title{\bf Rare processes in ultrahigh-energy tau-lepton transport
	
}

\author{Guo-yuan Huang}
\email{huangguoyuan@cug.edu.cn}
\affiliation{School of Mathematics and Physics, China University of Geosciences, 430074 Wuhan, China}

\date{\today}

\fontdimen1\font=0.0em
\fontdimen2\font=0.38em
\fontdimen3\font=0.2em
\fontdimen4\font=0.05em

\begin{abstract}
	\noindent In cosmic neutrino observatories, charged-lepton transport is a key input for interpreting observables and reconstructing neutrino events. Charged leptons propagating through matter undergo several energy loss processes, such as electron pair production and photonuclear interaction. In this work, we investigate several rare processes in tau-lepton transport, focusing primarily on muon pair production, $\tau N \to \tau \mu^+\mu^- N$, and Primakoff neutral-pion production, $\tau N \to \tau \pi^0 N$. 
	At EeV energies, we find that the energy loss contributions from muon pair and Primakoff pion production are only $0.6\%$ and $0.2\%$, respectively, of that from electron pair production. 
	Nevertheless, dimuon production may be relevant to tau neutrino searches in underwater and under-ice Cherenkov telescopes. The interaction length for dimuon emission by an EeV tau is only $6~{\rm km}$ in standard rock. Such events may be identified through the lateral separation of dimuon tracks or through a  kebab topology if tau decays in the detector.
\end{abstract}

\maketitle


\section{Introduction}

Over the past decades, IceCube observations have opened a new window into the high-energy universe and begun to shed light on the origins of ultrahigh-energy (UHE) neutrinos~\cite{IceCube:2013low,IceCube:2013cdw,IceCube:2018cha,IceCube:2022der,Abbasi:2023bvn,IceCube:2023ame,IceCube:2024nhk,IceCube:2024fxo,IceCubeCollaborationSS:2025jbi}. 
More recently, observations from KM3NeT have started to extend our exploration of the neutrino flux toward the EeV energy frontier~\cite{KM3NeT:2025npi}.	
A central goal of next-generation UHE neutrino telescopes is to identify the astrophysical sources of these neutrinos and characterize the diffuse neutrino flux with greater precision~\cite{KM3Net:2016zxf,TRIDENT:2022hql,Huang:2023mzt,Zhang:2024slv,Baikal-GVD:2022fis,IceCube-Gen2:2020qha,RNO-G:2020rmc,Neronov:2016zou,GRAND:2018iaj,Otte:2018uxj,Otte:2019aaf,ARA:2019wcf,Abarr:2020bjd,Anker:2020lre,POEMMA:2020ykm,TAMBO:2025jio,Wissel:2020fav,Wissel:2020sec,Ogawa:2021dK,deVries:2021BA}. Achieving this goal requires not only larger instrumented volumes but also a detailed understanding of how neutrinos and their secondary particles interact with matter.

UHE neutrinos are observed primarily through the charged-current interaction, in which a neutrino converts into a charged lepton via $W$-boson exchange. Depending on the lepton flavor, the event may appear as a cascade or a track in underwater or under-ice detectors~\cite{Huang:2023mzt}.
For neutrino telescopes that observe air showers (such as GRAND~\cite{GRAND:2018iaj}, POEMMA~\cite{POEMMA:2020ykm}, BEACON~\cite{Wissel:2020fav}, TAMBO~\cite{TAMBO:2025jio} et al.), the dominant signal is the decay of a tau lepton that emerges from a mountain or the ground after propagating tens of kilometers~\cite{Berezinsky:1975zz,Domokos:1997ve,Domokos:1998hz,Capelle:1998zz,Fargion:1999se,Fargion:2000iz,LetessierSelvon:2000kk,Feng:2001ue,Kusenko:2001gj,Bertou:2001vm,Cao:2004sd,Zas:2005zz,Baret:2011zz}.
There are even more ambitious proposals that aim to detect UHE neutrinos (and cosmic rays) through the Moon, such as SKA-Low~\cite{Huege:2026beb}, by measuring the Askaryan radio signals generated by their interactions with the lunar regolith.
A reliable description of charged-lepton propagation in matter is therefore essential for accurately reconstructing the energy and direction of the parent neutrino.

The charged-lepton transport is governed by several energy loss processes in matter, including bremsstrahlung, electron pair production, ionization, and photonuclear interactions~\cite{Tsai:1973py,Sokalski:2000nb,koehne2013proposal,Safa:2021ghs,Alameddine:2023wrp}.
These processes determine how far a charged lepton can travel from the interaction vertex. The same energy loss processes  also deposit energy along its trajectory, producing Cherenkov light that makes the track visible.

Above $100~{\rm TeV}$, the dominant energy loss channels in muon and tau transports are the photonuclear interaction, and the electron pair production $\ell N \to \ell e^+ e^- N$~\cite{koehne2013proposal}, where $\ell$ denotes the charged lepton and $N$ the nuclear target. At sufficiently high lepton energies, the required momentum transfer $q$ from the nucleus in the $t$-channel photon exchange can be very small, such that the cross section can be significantly enhanced.
Beside those leading channels, several rare processes in muon propagation have also been studied, including $\mu\, N \to \mu \, \mu^+ \mu^- N$~\cite{Kudryavtsev:1999zu,Kelner:2000va,Maciuc:2006xb,Berezhnoy:2007uz,Alameddine:2023wrp,Abbiendi:2024swt} and $\mu\, N \to \mu \, \tau^+ \tau^- N$~\cite{Akhundov:1979bd,Akhundov:1979wp,Bulmahn:2008fa,Bulmahn:2010qna,Bulmahn:2010pg,Das:2026eyy}. Their contributions to energy loss are found to be small and can often be neglected in numerical simulations of lepton transport~\cite{Alameddine:2023wrp}. Nevertheless, they can produce distinctive signatures in underwater and under-ice neutrino telescopes~\cite{Kudryavtsev:1999zu,Kelner:2000va,Bulmahn:2008fa,Bulmahn:2010qna}. For example, muon pair production during muon transport can produce muon bundles~\cite{Kudryavtsev:1999zu,Kelner:2000va}, whereas tau pair production can yield ditau-bang signatures~\cite{Bulmahn:2008fa,Bulmahn:2010qna,Huang:2022ebg}.
The conversion of $\mu$ into $\nu_\mu$ through weak interactions has also been considered a rare process~\cite{Alameddine:2023wrp}, but it can contribute a background to the lollipop signal in tau neutrino searches.

In this work, we focus on tau-lepton propagation, which has received less attention than muon propagation despite its growing importance. Tau transport is fundamental to tau neutrino detection and may also help underwater and under-ice telescopes isolate the tau-flavor component of the neutrino flux. Here, we mainly investigate two unexplored  channels and discuss their possible implications for future tau neutrino searches: (i) muon pair production, $\tau N \to \tau \mu^+\mu^- N$, and (ii) Primakoff pion production, $\tau N \to \tau \pi^0 N$. The corresponding Feynman diagrams at the tree level are shown in Fig.~\ref{fig:Feyn}.
Both processes become increasingly important as the tau energy increases. This behavior can be understood within the equivalent-photon approximation (EPA)~\cite{vonWeizsacker:1934nji,Williams:1934ad}. At very high lepton energies, collinear photon emission is enhanced, with a distribution that scales as $\rho^{}_{\gamma} \sim \ln(E^{}_{\ell}/m^{}_{\ell})$. The process may then be viewed as the conversion of this photon into a muon pair or a neutral pion in the electromagnetic field of the nucleus.

The remainder of this work is organized as follows. In Sec.~\ref{sec:II}, we revisit the derivation of  cross sections for dimuon and pion productions and correct a coefficient in an expression of previous work. In Sec.~\ref{sec:III}, we calculate the energy loss contributions of these processes, and discuss their potential observable signatures. We also comment on the implication of $\tau \to \nu^{}_{\tau}$ due to weak interactions. We conclude in Sec.~\ref{sec:IV}.

\section{Cross sections} \label{sec:II}

\begin{figure}[t!]
	\begin{center}
		\includegraphics[width=0.5\textwidth]{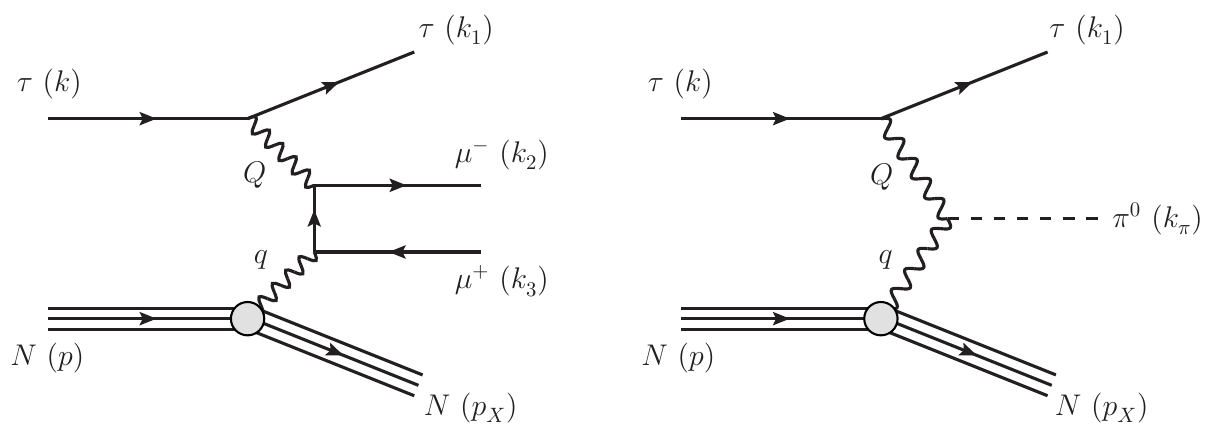}
	\end{center}
	\caption{Feynman diagrams for dimuon production (left) and Primakoff neutral-pion production (right) during tau-lepton propagation in matter. For dimuon production, an additional diagram with $\mu^+$ and $\mu^-$ exchanged is not shown. The momentum is indicated along the diagram lines.}
	\label{fig:Feyn}
\end{figure}

For the processes involving two $t$-channel exchanges shown in Fig.~\ref{fig:Feyn}, the cross section is dominated by the region of phase space with small momentum transfers. In particular, the minimum momentum transfer scales approximately as $q^{2}_{\rm min} \approx m^{2}_{\tau} m^2_{V}/E^{2}_{\tau}$~\cite{Tsai:1973py}, where $m^{}_{V} = 2 m^{}_{\mu}$ or $m^{}_{\pi}$ is the total mass of the newly produced state.
For $E^{}_{\tau} = 100~{\rm TeV}$, dimuon production has $q^{}_{\rm min} \sim 3.7~{\rm keV}$, corresponding to a length scale comparable to the atomic size.
The cross section at ultrahigh energies is therefore dominated by inverse momentum transfers larger than the nuclear size, so the scattering benefits from a coherent $Z^2$ enhancement, but smaller than the atomic size, where screening becomes important. 
%


%
In principle, calculating cross sections for the pair production requires direct integration over the four-body final-state phase space~\cite{Bulmahn:2008fa}. Although the EPA and other approximations can simplify the cross section calculation~\cite{Ivanov:1998dv}, deriving parameterized expressions based on these approximations is beyond the scope of this work.
Moreover, the accuracy of the EPA is subject to the photon virtuality, which must not be too large. We therefore choose to directly integrate the $2\to 4$ phase space for coherent nuclear scatterings.

\subsection{Cross section for muon pair production}
%
The computational framework developed in Refs.~\cite{Akhundov:1979bd,Akhundov:1979wp,Bulmahn:2008fa} can readily be applied to muon pair production.
We follow Ref.~\cite{Bulmahn:2008fa}, which provides a detailed description and adopts a more modern convention. However, we find a factor-of-four discrepancy in the formalism of the cross section presented therein. To clarify the origin of this difference, we elaborate on the relevant derivation below, following the notation of Ref.~\cite{Bulmahn:2008fa} wherever possible, and compare the results at the points where they differ.

For the muon pair production,
the spin-averaged squared matrix element can be represented by
\begin{align}
	\frac{1}{4}\sum_{\rm spins}^{}|\mathcal{M}|^2 = e^8 A^{}_{\alpha \beta}\cdot B^{\alpha\beta}_{\mu\nu}\cdot 4\pi W^{\mu\nu}_{} \cdot \frac{1}{t^2 Y^2}  \;,
\end{align}
where $t \equiv -q^2$ and $Y \equiv -Q^2$ are the momentum transfers associated with the nucleus and the charged lepton, respectively. The tensor $A^{}_{\alpha\beta}$ denotes the contribution from the tau-lepton current, and $B^{\alpha\beta}_{\mu\nu}$ is the corresponding tensor for the lepton pair. The hadronic tensor is $W^{\mu\nu}_{} = -g^{\mu\nu}F^{}_{1} + 2 p^\mu p^\nu F^{}_{2} / t$, where the structure functions $F^{}_{1}$ and $F^{}_{2}$ encode the nuclear form factor and the atomic screening effect~\cite{Bulmahn:2008fa}.

The cross section can be expressed as
\begin{align} \label{eq:dSigma}
	\mathrm{d}\sigma = \frac{\alpha^4}{\pi^4\sqrt{\lambda^{}_{S}}}  \frac{A^{}_{\alpha \beta}\cdot B^{\alpha\beta}_{\mu\nu}\cdot  W^{\mu\nu}_{}}{t^2 Y^2}  \mathrm{d}{\Phi^{}_{\rm PS}}\;,
\end{align}
where $\lambda^{}_{S} \equiv (2 p \cdot k)^2 - 4 m^2_\tau M^{2}_{N}$ and $M^{}_{N}$ is the mass of the target nucleus.
The original expression for the phase space in Ref.~\cite{Bulmahn:2008fa} is not explicitly provided. Therefore, we specify our choice of phase space below,
\begin{align} \label{eq:dPhi}
	\mathrm{d}{\Phi^{}_{\rm PS}} =	
	{\mathrm{d}^4 q}  \frac{\mathrm{d}^3 k^{}_{1}}{2E^{}_{1}} \,
	\frac{\mathrm{d}^3 k^{}_{2}}{2E^{}_{2}} 
	\frac{\mathrm{d}^3 k^{}_{3}}{2E^{}_{3}} 
	\delta^4 (k+q-\sum^{3}_{i=1} k^{}_{i}) \;.
\end{align}
To perform the integration, we change to a set of variables suited to the two $t$-channel exchanges. Following Ref.~\cite{Bulmahn:2008fa}, we introduce $(S^{}_{x}, Y, M^2_{x}, t, V^2, \phi^{}_{1}, \phi^{}_{q})$, with
\begin{align}
	&	S^{}_{x}  \equiv 2 p \cdot Q \;, \hspace{0.3cm}
	Y  \equiv -Q^2 \notag\;, \hspace{0.3cm}
	M^{2}_{x}  \equiv (p - q)^2 \notag\;,\\
	&	t  \equiv  -q^2 \;,  \hspace{0.3cm}
	V^2  \equiv (q+Q)^2 \;,
\end{align}
where $\phi^{}_{1}$ ($\phi^{}_{q}$) is the azimuth angle of $\bm{k}^{}_{1}$ ($\bm{q}$), with $\bm{k}$ ($\bm{Q}$) chosen as the $z$ axis in the rest frame of the target nucleus.

The first two differential elements ${\mathrm{d}^4 q}\,{\mathrm{d}^3 k^{}_{1}}$ in Eq.~(\ref{eq:dPhi}) can be written as
\begin{align}
	\frac{\mathrm{d}^3 \bm{k}^{}_{1}}{2E^{}_{1}}   & = \frac{1}{4\sqrt{\lambda^{}_{S}}}\mathrm{d} S^{}_{x} \mathrm{d} Y \mathrm{d}\phi^{}_{1} \;,\\
	{\mathrm{d}^4 q}  & = \frac{1}{4\sqrt{\lambda^{}_{Y}}} \mathrm{d} M^2_x  \mathrm{d} t \mathrm{d} V^2 \mathrm{d} \phi^{}_{q} \;,
\end{align}
where $\lambda^{}_{Y} \equiv S^2_x + 4 M^2_N\, Y$. In deriving the above relations, we have used the Jacobians 
\begin{small}
	\begin{align}
		\frac{\partial (S^{}_{x}, Y)}{ \partial (|\bm{k}^{}_{1}|, \cos{\theta^{}_{1}})} & = 
		\left| \begin{matrix}
			{-2M^{}_{t} |\bm{k}^{}_{1}|}/{E^{}_{1}} & 0 \vspace{4pt}\\
			{2 E^{}_{k} |\bm{k}^{}_{1}|}/{ E^{}_{1}} - 2 |\bm{k}| \cos{\theta}&  - 2 |\bm{k}| |\bm{k}^{}_{1}|
		\end{matrix}\right| \notag\\[5pt] & = \frac{2\sqrt{\lambda^{}_{S}} |\bm{k}^{}_{1}|^2}{E^{}_{1}}\;, \\[5pt]
		\frac{\partial (M^{2}_{x}, t, V^2)}{ \partial (E^{}_{q}, |\bm{q}|, \cos{\theta^{}_{q}})} & = \left| \begin{matrix}
			-2 M^{}_{t} & -2|\bm{Q}| & 0  \vspace{4pt}\\
			2 E^{}_{q} & -2|\bm{q}| & 0  \vspace{4pt}\\
			2E^{}_{Q} + 2 E^{}_{q} & 2 |\bm{q}|  - 2 |\bm{Q}| \cos{\theta}& -2|\bm{q}| |\bm{Q}|
		\end{matrix}\right|\notag \\[5pt]  & = 4 \sqrt{\lambda^{}_{Y}} |\bm{q}|^2\;.
	\end{align}
\end{small}
Now, the phase space in Eq.~(\ref{eq:dPhi}) can be recast as
\begin{align}
	\mathrm{d}{\Phi^{}_{\rm PS}} =	
	\frac{1}{16\sqrt{\lambda^{}_{S} \lambda^{}_{Y}}}\mathrm{d} S^{}_{x} \mathrm{d} Y \mathrm{d}\phi^{}_{1} \, \mathrm{d} M^2_x  \mathrm{d} t \mathrm{d} V^2 \mathrm{d} \phi^{}_{q} \,	\mathrm{d}{\rm \Phi^{}_{\rm pair}}\;,
\end{align}
which agrees with the final expression of phase space  used in the appendix of Ref.~\cite{Bulmahn:2008fa}. Here, $\mathrm{d}{\rm \Phi^{}_{\rm pair}}$ is the phase space of the outgoing lepton pair,
\begin{align}
	\mathrm{d}{\rm \Phi^{}_{\rm pair}} & =	
	\frac{\mathrm{d}^3 k^{}_{2}}{2E^{}_{2}} 
	\frac{\mathrm{d}^3 k^{}_{3}}{2E^{}_{3}} 
	\delta^4 (k+q-\sum^{3}_{i=1} k^{}_{i})  \notag\\
	& = \frac{1}{8} \sqrt{1- \frac{4 m^2_e}{V^2}}\, \mathrm{d}\mathrm{cos}\theta^{}_{e}  \mathrm{d}\phi^{}_{e}\;,
\end{align}
where $\theta^{}_{e}$ and $\phi^{}_{e}$ are the zenith and azimuth angles of $\bm{k}^{}_{2}$ in the center-of-mass frame of the pair.

While the final expressions for the phase space agree, our differential cross section in Eq.~(\ref{eq:dSigma}) now differs from Eq.~(A1) of Ref.~\cite{Bulmahn:2008fa} by a factor of four.
To determine whether this discrepancy is merely typographical, we try to reproduce the numerical result reported there.
We calculate the cross section for $\mu A \to \mu e^+ e^- X$ in standard rock, taking $Z = 11$ and $A = 22$. The result in Fig.~3 of Ref.~\cite{Bulmahn:2008fa} gives a total coherent-scattering cross section of $\sigma \approx 6.5 \times 10^{-25}~{\rm cm}^2$ at $E^{}_{\mu} = 10^8~{\rm GeV}$. Our integration yields $\sigma \approx 6.45 \times 10^{-25}~{\rm cm}^2$, which is in agreement with their result within the $0.1\%$ numerical precision.

As a benchmark for the tau transport, we  calculate the cross section for $\tau N \to \tau \mu^+\mu^- N$ in standard rock. At $E^{}_{\tau} = 10^9~{\rm GeV}$, we obtain $\sigma \approx 1.8 \times 10^{-29}~{\rm cm^2}$. This is only $3\times 10^{-5}$ of the cross section for electron pair production from tau. 
However, as we will see later, the corresponding contribution to energy loss is much larger than this cross-section ratio would suggest.

\subsection{Cross section for Primakoff neutral-pion production}

The calculation of the Primakoff process induced by charged leptons $\tau N \to \tau \pi^0 N$ is similar to that of pair production. The framework developed in the previous subsection can apply with minor modifications to the matrix element and the phase space.

The neutral pion couples to two photons through the axial anomaly. In the chiral limit, the amplitude for the $\pi \gamma \gamma$ coupling vertex is given by~\cite{Roberts:1994hh,Frank:1994gc,Maris:2002mz}
\begin{align}
	\Lambda^{\pi^0\gamma\gamma}_{\mu\nu}(q, Q) = \mathrm{i}\frac{  \alpha}{\pi f^{}_{\pi}} \epsilon^{}_{\mu\nu\rho\sigma} q^{\rho}_{}Q^{\sigma}_{} \;,
\end{align}
where $f^{}_{\pi} \approx 92~{\rm MeV}$ is the pion decay constant and $\epsilon$ is the Levi-Civita symbol. Because the process is dominated by small momentum transfers, we use this low-energy amplitude without introducing a momentum-dependent form factor. The resulting matrix element tensor is
\begin{align}
	B^{\alpha\beta}_{\mu\nu}  &=  \epsilon^{\ \alpha \rho\sigma}_{\mu} \epsilon^{\ \beta\rho^\prime \sigma^\prime}_{\nu} q^{}_{\sigma}Q^{}_{\rho} q^{}_{\sigma^\prime} Q^{}_{\rho^\prime} \;.
\end{align}
The corresponding differential cross section reads
\begin{align}
	\mathrm{d}\sigma = \frac{\alpha^4}{2\pi^5 f^2_\pi\sqrt{\lambda^{}_{S}}}  \frac{A^{}_{\alpha \beta}\cdot B^{\alpha\beta}_{\mu\nu}\cdot  W^{\mu\nu}_{}}{t^2 Y^2}  \mathrm{d}{\Phi^{}_{\rm PS}}\;.
\end{align}

The phase space is simpler than that for pair production and is given by
\begin{align}
	\mathrm{d}{\Phi^{}_{\rm PS}} & =	
	{\mathrm{d}^4 q}  \frac{\mathrm{d}^3 k^{}_{1}}{2E^{}_{1}} 
	\frac{\mathrm{d}^3 k^{}_{\pi}}{2E^{}_{\pi}} 
	\delta^4 (k+q- k^{}_{1} - k^{}_{\pi}) \notag\\
	& = \frac{1}{16\sqrt{\lambda^{}_{S} \lambda^{}_{Y}}}\mathrm{d} S^{}_{x} \mathrm{d} Y \mathrm{d}\phi^{}_{1} \, \mathrm{d} M^2_x  \mathrm{d} t  \mathrm{d} \phi^{}_{q} \;,
\end{align}
with $V^2 = m^2_{\pi}$. This expression can then be integrated to obtain the cross section.

\section{Observable effects} \label{sec:III}

Those new processes will contribute to tau energy loss, although they are expected to be subdominant compared with the standard channels. Nevertheless, quantifying their relative contributions is essential for a complete understanding of tau propagation. Furthermore, these subleading channels may produce distinctive observational signatures that could provide valuable probes of their underlying physics.

\begin{figure}[t!]
	\begin{center}
		\includegraphics[width=0.43\textwidth]{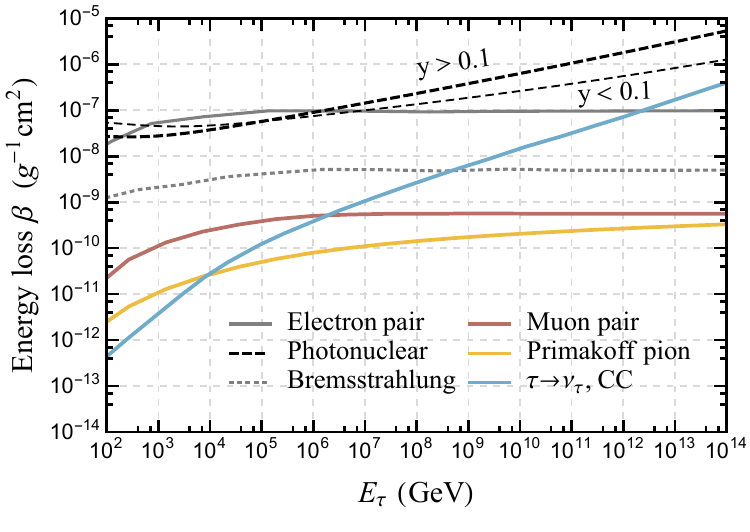}
	\end{center}
	\caption{Energy loss parameter $\beta$ as a function of the tau energy $E^{}_{\tau}$. Muon pair production and Primakoff pion production are shown by the solid red and yellow curves, respectively. The standard contributions from electron pair production, photonuclear reaction (ALLM model), and bremsstrahlung are included for comparison~\cite{koehne2013proposal}. We also present the energy loss due to $\tau \to \nu^{}_{\tau}$ induced by the charged-current interaction~\cite{Gandhi:1998ri}, which has been considered as a rare process for charged leptons in the literature. For the photonuclear interaction, we isolate  contributions from different energy loss fractions, namely $y > 0.1$ and $y < 0.1$. }
	\label{fig:loss}
\end{figure}

\begin{figure}[t!]
	\begin{center}
		\includegraphics[width=0.43\textwidth]{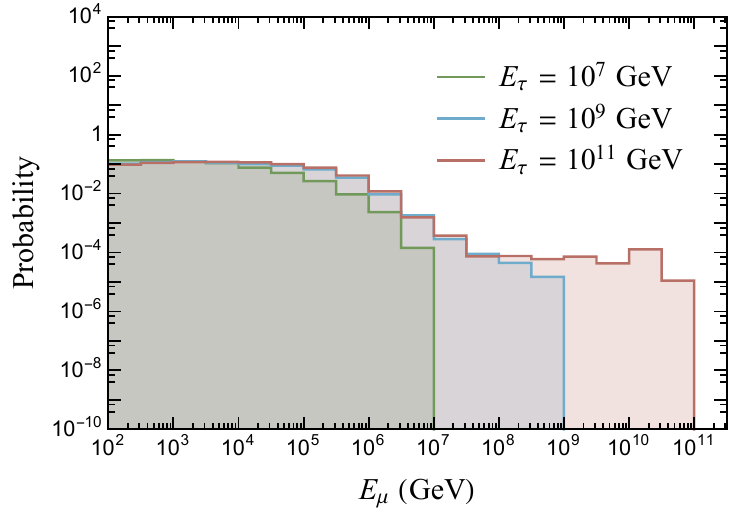}
		\includegraphics[width=0.432\textwidth]{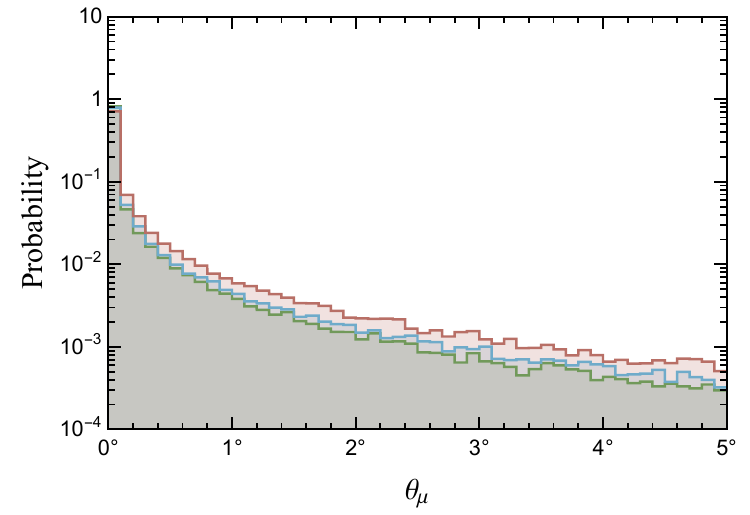}
	\end{center}
	\caption{Normalized distributions of the energy of the emitted muons (upper panel) and the angle between one of the muons and the initial tau (lower panel). The initial tau energies are chosen as $E^{}_{\tau} = 10^7~{\rm GeV}$ (in green), $10^9~{\rm GeV}$ (in blue), and $10^{11}~{\rm GeV}$ (in red).}
	\label{fig:ProbEmuTheta}
\end{figure}

\subsection{Energy loss contributions}

Neglecting tau decay, the evolution of the tau energy in matter can be parameterized as $\mathrm{d}E/\mathrm{d}X = -\alpha -\beta E$, where $\alpha$ and $\beta$ characterize the average energy loss rate and depend only weakly on the energy $E$. The parameter $\alpha$ accounts for ionization, whereas $\beta$ receives its main contributions from pair production, bremsstrahlung, and photonuclear interactions. The parameter $\beta$ is calculated as
\begin{align}
	\beta = \int^{}_{} \mathrm{d}y \, y\, \frac{\mathrm{d}\sigma}{\mathrm{d} y} \cdot \frac{N^{}_{\rm A}}{A} \;,
\end{align}
where $y$ is the fractional energy loss per collision, $N^{}_{\rm A}$ is Avogadro's constant, and $A$ is the mass number of the target atom.

Fig.~\ref{fig:loss} shows the energy loss parameter $\beta$ as a function of the tau energy $E^{}_{\tau}$ for several processes, taking water/ice as the medium.
The contributions from electron pair production (solid black curve) and bremsstrahlung (dotted black curve) approach their asymptotic values above $E^{}_{\tau} = 10^6~{\rm GeV}$.
The red and yellow curves represent the contributions from muon pair and Primakoff pion production, respectively. At $E^{}_{\tau} = 10^9~{\rm GeV}$, the muon pair contribution is approximately $0.6\%$ of the electron pair contribution, even though its cross section is only $0.003\%$ as large. This difference arises because the cross section  for electron pair production is dominated by the region of phase space with very small momentum transfers and consequently very small energy loss fractions $y$.
Although the neutral-pion contribution continues to increase across the energy window shown in the plot, at $E^{}_{\tau} = 10^9~{\rm GeV}$ it is only $0.2\%$ of the electron pair contribution. Given the expected precision of future neutrino telescopes~\cite{KM3Net:2016zxf,TRIDENT:2022hql,Huang:2023mzt,Zhang:2024slv,Baikal-GVD:2022fis,IceCube-Gen2:2020qha,RNO-G:2020rmc,Neronov:2016zou,GRAND:2018iaj,Otte:2018uxj,Otte:2019aaf,ARA:2019wcf,Abarr:2020bjd,Anker:2020lre,POEMMA:2020ykm,TAMBO:2025jio,Wissel:2020fav,Wissel:2020sec,Ogawa:2021dK,deVries:2021BA} and the uncertainties in the standard energy loss processes~\cite{koehne2013proposal,Safa:2021ghs,Alameddine:2023wrp}, those two channels should safely be neglected in lepton-transport simulations.

In Fig.~\ref{fig:loss}, we have also shown the energy loss contribution from the conversion of $\tau$ into $\nu_{\tau}$ through $W$-boson exchange in weak interactions (blue curve). 
The weak energy loss is obtained with 
\begin{align}
\beta^{}_{\rm weak} = \sigma^{}_{\rm CC} N^{}_{A} \;,
\end{align}
where the charged-current cross section $\sigma^{}_{\rm CC}$ is taken from Ref.~\cite{Gandhi:1998ri}.
This process has traditionally been considered extremely rare and negligible compared with photonuclear and pair production~\cite{Alameddine:2023wrp}. As shown in the figure, this assumption remains valid at lower tau energies. 
For $E_{\tau} \gtrsim 10^{10}~\mathrm{GeV}$, the charged-current scattering will dominate the conversion of $\tau \to \nu^{}_{\tau}$ over the tau decay.
For $E_{\tau} \gtrsim 10^{12}~\mathrm{GeV}$ (1~ZeV), the weak interaction surpasses electron pair production, although its contribution remains smaller than that of photonuclear interactions (dashed black curves). We also note that, at $1~{\rm ZeV}$, the energy loss due to photonuclear interactions is dominated by the hard component with inelasticity $y > 0.1$. Here, we want to comment on its implications for neutrino searches at ZeV energies.

While most existing and proposed neutrino telescopes are not sensitive to ZeV neutrinos, this energy range is the primary target of lunar detection concepts for UHE cosmic neutrinos. Among these proposals, the SKA-Low telescope is expected to provide one of the most promising sensitivities~\cite{Bray:2015ota}. SKA-Low aims to detect radio pulses generated by cascades induced by cosmic particles in the lunar regolith via the Askaryan effect. The coherent radio emission from these cascades can be measured by the low-frequency antenna array, enabling observations of neutrinos at energies far beyond the reach of conventional detectors.

SKA-Low is sensitive to all neutrino flavors. For $\nu_{\tau}$, however, previous sensitivity studies have typically discarded the produced tau lepton, assuming that only the initial interaction contributes to the radio signal~\cite{James:2008ff,Bray:2015ota}. In this treatment, approximately $20\%$ of the incident neutrino energy is deposited in the hadronic cascade at the interaction vertex, while the remaining $80\%$ carried away by $\tau$ has been neglected considering  $\tau$'s soft energy loss along its track.
This approximation should break down when hadronic cascades from hard photonuclear interactions, i.e., $\tau N \to \tau X$ via a photon exchange, are taken into account. During propagation, the secondary tau lepton can repeatedly initiate hadronic cascades through photonuclear interactions, producing Askaryan radio emissions as first noted in Ref.~\cite{Garcia-Fernandez:2020dhb}.
The contributions to the energy loss from different inelasticity ranges are illustrated by two dashed curves in Fig.~\ref{fig:loss}, representing the soft component ($y<0.1$), and the hard component ($y>0.1$). It can be noticed that hard energy losses become increasingly important at higher tau energies. At $E_{\tau}=10^{14}~\mathrm{GeV}$, photonuclear interactions with inelasticity $y>0.1$ account for approximately $80\%$ of the total energy loss. As a consequence, each such interaction can deposit an energy comparable to that of the initial neutrino interaction vertex. Therefore, it is crucial to include these stochastic cascades induced by secondary taus in sensitivity estimates for lunar neutrino searches.

\subsection{Dimuon signal from tau}

Dimuon production during high-energy muon transport has been studied by previous works~\cite{Kudryavtsev:1999zu,Kelner:2000va} using simplified parameterizations of the cross section. Similar dimuon signatures can also arise from other Standard Model (SM) processes.
A well-known source is the charm production in charged-current deep inelastic scattering,
$\nu^{}_\mu + N \to \mu + {\rm c} + X$ followed by $ {\rm c} \to {\rm s}+\mu + \nu^{}_\mu$~\cite{Zhou:2021xuh,Sun:2022lti,Ternes:2024lsj}.
Electroweak $W$-boson and trident production channels provide additional SM contributions~\cite{Ge:2017poy,Zhou:2019frk,Zhou:2019vxt}. 
Anomalous double-track signatures may also arise in new physics scenarios involving heavy neutral leptons~\cite{Meade:2009mu}, secluded dark matter~\cite{ANTARES:2016obx}, leptogluons~\cite{Bai:2025pef}, supersymmetric particles~\cite{Albuquerque:2003mi,Ahlers:2006pf}, boosted dark matter~\cite{Dev:2025czz} and so on.
The typical signature of dimuon events consists of two tracks registered simultaneously in the detector, provided that the lateral separation between these two tracks is resolvable. Otherwise, they will be reconstructed just as a single track.

Likewise, dimuons can be produced along the trajectories of energetic tau leptons. 
We find that such production is quite frequent for an EeV tau. 
For muon transport, it is challenging to identify dimuons, as the resulting dimuon system is highly boosted and buried within the signal of the parent muon track.
The tau lepton, however, provides a unique topology.
In particular, tau track can naturally end up with a lollipop event when the tau decays inside the detector. In addition to laterally separated tracks, the accompanying high-energy dimuons continue to propagate beyond the tau-decay vertex, potentially giving rise to a distinctive signature.
In the following, we will discuss the lateral separation and the potential new signature one by one.

\subsubsection{Lateral separation}
For track events in underwater or under-ice telescopes, the interaction vertex may lie either inside or outside the instrumented volume, corresponding to starting and through-going tracks, respectively.
The dimuon track event has a distinctive topology with two energetic muons correlated in time and direction.
Because  two muons are strongly boosted along the direction of the incident particle, 
the  identification of such events is fundamentally limited by the detector's capability to  resolve the lateral separation $D_{\mu\mu}$ between them~\cite{Zhou:2021xuh}.
In IceCube, the large instrumented volume is achieved using a sparse array of digital optical modules (DOMs). The horizontal (or vertical) separation between DOMs is approximately $125~{\rm m}$ (or $17~{\rm m}$), which sets a  lower bound on the lateral separation resolution. Consequently, a rough threshold of 34 meters is typically required to effectively distinguish two parallel muon tracks~\cite{Zhou:2021xuh,Sun:2022lti}. Because of this wide spacing, searches for dimuon events in IceCube are generally restricted. 
Future neutrino telescopes, such as KM3NeT~\cite{KM3Net:2016zxf}, are expected to improve the dimuon sensitivity through denser detector configurations. Simulations indicate that  arrays like KM3NeT can resolve dimuon tracks with lateral separations down to the order of just a few meters~\cite{Sun:2022lti}.

\begin{figure}[t!]
	\begin{center}
		\includegraphics[width=0.38\textwidth]{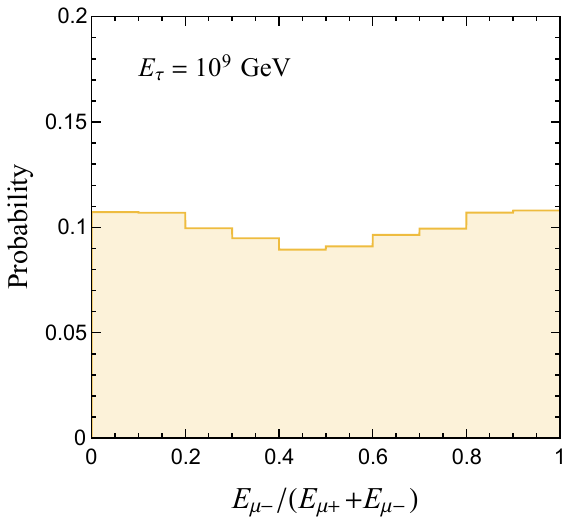}
	\end{center}
	\caption{The normalized  distribution of the energy ratio defined by $E^{}_{\mu^-}/(E^{}_{\mu^+} + E^{}_{\mu^-})$.}
	\label{fig:ERatioAngle}
\end{figure}

For tau-lepton transport, to assess whether the two produced tracks can be resolved through their lateral separation, we examine the kinematic properties of dimuons. The upper panel of Fig.~\ref{fig:ProbEmuTheta} shows the normalized distribution of muon energy for initial tau energies $E^{}_{\tau} = 10^7~{\rm GeV}$, $10^9~{\rm GeV}$, and $10^{11}~{\rm GeV}$, respectively. Fig.~\ref{fig:ERatioAngle} illustrates that the pair energy is shared  evenly between  two muons. Because the energy distributions of  two muons are symmetric, we do not distinguish between them. As shown in Fig.~\ref{fig:ProbEmuTheta}, an individual muon can carry an energy of the same order of magnitude as the initial tau, but the probability of such an event is only about $10^{-5}$--$10^{-4}$.  For $E^{}_{\tau} = 1~{\rm EeV}$, the average muon energy is about $1.3~{\rm TeV}$, which is well above the threshold of a typical water Cherenkov telescope.

Here, two muons are also nearly collinear with the initial tau lepton. The lower panel of Fig.~\ref{fig:ProbEmuTheta} shows the distribution of the angle $\theta^{}_{\mu}$ between the outgoing muon and the initial tau for several tau energies. The distribution depends only weakly on energy. At $E^{}_{\tau} = 10^9~{\rm GeV}$, the probability that this angle exceeds $0.1^{\circ}$ ($0.05^{\circ}$) is approximately $20\%$ ($30\%$).
If the two muons are emitted with a nonzero opening angle, their separation increases as they propagate. For an angle $\theta_{\mu}$ and propagation distance $L$, the characteristic lateral displacement is approximately
$d_\perp \sim L \, \theta_{\mu}$.
For a resolvable separation of $5~{\rm m}$ and a propagation distance of $\mathcal{O}(5~{\rm km})$, the corresponding minimum angle is approximately $\mathcal{O}(0.06^{\circ})$.

If  two muon tracks overlap completely, geometric discrimination is no longer possible. In principle, they may still be distinguished statistically through their energy loss patterns~\cite{Kistler:2016ask}. But in practice, the stochastic nature of muon energy loss makes this discrimination approach very challenging.

\subsubsection{Kebab events}
In underwater and under-ice Cherenkov telescopes, the identification of tau neutrinos relies heavily on distinct event topologies with low backgrounds, such as double bangs and lollipops~\cite{Beacom:2003nh,Cowen:2007ny}. 
While the classic double bang topology is a rather clean signature~\cite{Learned:1994wg}, it is constrained by the physical size of the detector. 
The  tau decay length scales approximately as $L^{}_\tau \approx 50~\mathrm{m} \times (E^{}_\tau / 1~\mathrm{PeV})$. At PeV energies, the decay length is on the order of tens of meters, allowing both cascades of a double bang to be  resolved while contained within a cubic-kilometer detector like IceCube. However, as the energy approaches the EeV regime, $L_\tau$ grows to tens of kilometers, far exceeding the typical size of such instruments.
In this case, an alternative signature is the lollipop topology~\cite{Beacom:2003nh,IceCube:2012lak}. 
In a lollipop event, the initial $\nu_\tau$ interaction occurs outside the instrumented volume. After kilometer-scale transport, the resultant tau track ends up as a hadronic or electromagnetic cascade inside the detector.
Above $10~{\rm PeV}$, the lollipop topology significantly dominates the observable $\nu_\tau$ event rate~\cite{Beacom:2003nh}. 

\begin{figure}[t!]
	\begin{center}
		\includegraphics[width=0.4\textwidth]{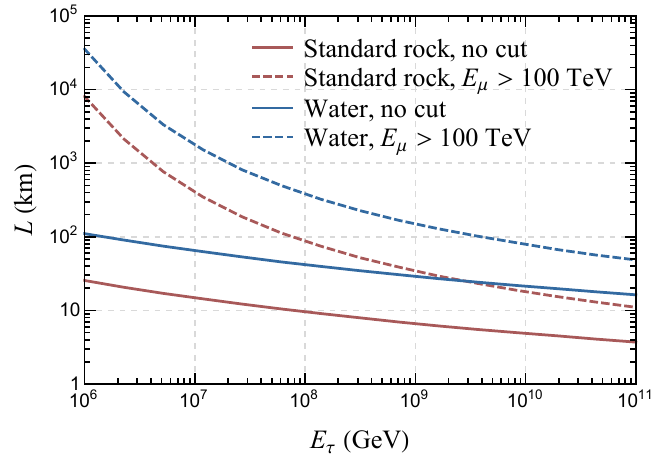}
	\end{center}
	\caption{The interaction length for muon pair production as a function of the tau energy $E^{}_{\tau}$, defined by $L\equiv (\sigma n)^{-1}$, where $\sigma$ is the cross section and $n \equiv \rho\, N^{}_{\rm A}/A$ is the number density of target nuclei. The solid curves show the inclusive result, whereas the dashed curves require at least one muon to have $E^{}_{\mu} > 100~{\rm TeV}$. Results are shown for standard rock (red) and water (blue).}
	\label{fig:intLength}
\end{figure}

Dimuon production may have implications for those lollipop events.
To assess, an important quantity is the number of dimuons produced along the trajectory of the tau. In Fig.~\ref{fig:intLength}, we show the interaction length, which is the average distance a particle travels before undergoing a  scattering event, as a function of the tau energy. At $E_{\tau} = 10^9~\mathrm{GeV}$, the tau travels an average distance of $L \approx 50~\mathrm{km}$ before decay. In comparison, the relevant interaction length for dimuon emission is approximately $L \approx 5~\mathrm{km}$ in standard rock. This implies that the EeV tau will generate roughly $\mathcal{O}(10)$ dimuon events during its propagation.

An immediate consequence is that the lollipop topology will potentially be accompanied by a muon pair. We can roughly estimate the probability of these muons reaching the detector alongside a  lollipop event. Assuming a detection energy threshold of $10~\mathrm{GeV}$ for muons and an energy cut at the production point of $E_{\mathrm{0}} = 100~\mathrm{TeV}$, the maximum distance a muon can travel is determined by~\cite{Kelner:2000va}:
\begin{equation}
	E(X) = -\frac{\alpha}{\beta} + \left(\frac{\alpha}{\beta} + E_{0}\right) \mathrm{e}^{-\beta X} \;.
\end{equation}
For muons propagating in standard rock, adopting $\alpha = 2.23 \times 10^{-3}~{\rm GeV\cdot cm^2/g}$ and $\beta = 4.64 \times 10^{-6}~{\rm cm^2 /g}$~\cite{koehne2013proposal} yields a maximum propagation distance of $L \approx 4~\mathrm{km}$. 
As shown in Fig.~\ref{fig:intLength}, the average interaction length required to produce a muon with an energy above $100~\mathrm{TeV}$ is $40~\mathrm{km}$. Thus, approximately $10\%$ of such lollipop events will be accompanied by dimuons if the EeV tau is produced $4~\mathrm{km}$ away. For an initial tau energy of $10~\mathrm{EeV}$, this probability increases to $20\%$ and would become even larger at higher neutrino energies.

The background of concern for such signals originates right from tau-lepton decays. The muonic decay channel of the tau can produce a pure high-energy muon (without cascade association) and generate a sugardaddy-like event~\cite{Cowen:2007ny,Fagundes:2019wzy}, which should be distinguishable from lollipop with dimuon tracks.
However, the hadronic decay of the tau lepton (with a branching ratio about $65\%$) in the detector can generate low-energy muons via secondary meson decays within the cascade. If the energies of the muon pair are also low, they may be difficult to distinguish from the cascade muons.
A quantitative assessment of the influence would require a detailed shower simulation.

\section{Discussion} \label{sec:IV}

We have calculated the differential cross sections for rare processes during tau-lepton transport, including dimuon production and Primakoff neutral-pion production, and investigated their implications for UHE neutrino telescopes. At EeV energies, their contributions to tau energy loss are only $0.6\%$ and $0.2\%$, respectively, of the contribution from  electron pair production. They can therefore be neglected in transport simulations at the precision expected for current and next-generation neutrino telescopes. 
In addition, we also emphasize that hard photonuclear scatterings during tau transport become increasingly important at the energy scales relevant for lunar neutrino searches like SKA-Low.

Despite their small contributions to the energy loss, dimuons may produce distinctive signatures in underwater and under-ice Cherenkov detectors. At $E^{}_{\tau}=10^9~{\rm GeV}$, approximately $20\%$ of outgoing dimuons are emitted at an angle greater than $0.1^{\circ}$ relative to the initial tau direction, allowing the tracks to develop a potentially resolvable lateral separation. In addition, if the tau decays inside the detector, the companion muon pair can continue propagating through the detector, leaving tracks that extend beyond the tau decay vertex. The resulting topology appears more like a kebab than a traditional lollipop. 
We recognize that these dimuon events are very rare compared to the dominant experimental signals. Nevertheless, studying such events provides a complementary piece  to understand neutrino interactions and the propagation behavior of their associated charged leptons. 

\section*{Acknowledgments}
{\it The author is indebted to Yong Du for many inspiring discussions and for his warm hospitality during author's visit to the Institute of Modern Physics. 
The author also thanks Haoning He for stimulating discussions on the relevance of tau transport to SKA.
Feynman diagrams in Fig.~1 are generated by using Jaxodraw~\cite{Binosi:2008ig}.	
This work is supported by National SKA Program of China 2025SKA0110104 and by the ``CUG Scholar'' Scientific Research Funds at China University of Geosciences (Wuhan) under project No. 2024014.
}


\bibliographystyle{utcaps_mod}
\bibliography{reference}


\end{document}